\documentclass[12pt]{article}
\usepackage{graphicx}

\newcommand\pubnumber{SNSN-323-63}
\newcommand\pubdate{\today}

\def\napoli{LIP-Laborat\'orio de Instrumenta\c{c}\~ao e F\'\i sica Experimental de Part\'\i culas\\
Av. Elias Garcia, 1000-149 Lisboa, PORTUGAL}
\def\support{\footnote{Work supported by FCT (Funda\c{c}\~ao para a Ci\^encia e a Tecnologia), project CERN/FP/123600/2011.}}

\def\Title#1{\begin{center} {\Large #1 } \end{center}}
\def\Author#1{\begin{center}{ \sc #1} \end{center}}
\def\Address#1{\begin{center}{ \it #1} \end{center}}

\newcommand\pubblock{\rightline{\begin{tabular}{l} \pubnumber\\
         \pubdate  \end{tabular}}}
\newenvironment{Abstract}{\begin{quotation}  }{\end{quotation}}
\newenvironment{Presented}{\begin{quotation} \begin{center} 
             PRESENTED AT\end{center}\bigskip 
      \begin{center}\begin{large}}{\end{large}\end{center} \end{quotation}}

\def\beq{\begin{equation}}
\def\eeq#1{\label{#1}\end{equation}}
\def\eeqn{\end{equation}}

\def\beqa{\begin{eqnarray}}
\def\eeqa#1{\label{#1}\end{eqnarray}}
\def\eeqan{\end{eqnarray}}

\let\bar=\overbar

\def\Dslash{\not{\hbox{\kern-4pt $D$}}}
\def\dslash{\not{\hbox{\kern-2pt $\del$}}}

\def\msb{{\bar{\ssstyle M \kern -1pt S}}}

\begin{document}
\begin{titlepage}
\pubblock

\vfill
\Title{$\Delta$g/g results from the Open Charm production at COMPASS}
\vfill
\Author{Celso Franco\support, on behalf of the COMPASS collaboration}
\Address{\napoli}
\vfill
\begin{Abstract}
One of the main goals of the COMPASS experiment at CERN is the determination
of the gluon contribution to the nucleon spin. To achieve this goal COMPASS
uses a naturally polarised muon beam with an energy of 160 GeV and a fixed
polarised target. Two types of materials are used for the latter: $^{6}$LiD
(polarised deuterons) during the years of 2002-2006 and NH$_{3}$ (polarised
protons) in 2007. The gluons in the nucleon can be accessed directly via the
\textbf{P}hoton \textbf{G}luon \textbf{F}usion (PGF) process. Among the
channels studied by COMPASS, the production of charmed mesons is the one
that tags a PGF interaction in the most clean and efficient way. This talk
presents a result for the gluon polarisation, $\Delta g/g$, which is based
on a measurement of the spin asymmetry resulting from the production of
D$^{0}$ mesons. These mesons are reconstructed through the invariant mass of
their decay products. The statistical significance of the $D^{0}$ mass 
spectra has been improved significantly using a new method based on Neural 
Networks. The $\Delta g/g$ result is also presented using a next-to-leading 
order (NLO-QCD) analysis of the $\mu N \rightarrow q\bar{q}$ reaction. Such 
correction is relevant and was for the first time applied to an experimental 
measurement of the gluon polarisation.
\end{Abstract}
\vfill
\begin{Presented}
Charm 2012 - 5$^{\textrm{th}}$ International Workshop on Charm Physics\\
14-17 May 2012, Honolulu, Hawai'i 96822
\end{Presented}
\vfill
\end{titlepage}
\setcounter{footnote}{0}

\section{Introduction}

\indent It is known since the twenties that the nucleons carry half integer spin. However, what still remains unknown to the present day is the distribution of the $1/2$ spin among the nucleon constituents. Until the eighties it was believed that the nucleon spin was distributed among its three valence quarks. In fact, assuming an SU(2) flavour symmetry and an SU(2) spin symmetry we obtain the following expression for the static wave function of the proton (the spin projection is considered to be parallel to the quantisation axis):


\begin{equation}
|p\uparrow \rangle = \frac{1}{\sqrt{18}}\left[\ 2\ |u\uparrow u\uparrow d\downarrow \rangle\ - \ |u\uparrow u\downarrow d\uparrow \rangle \ - \ |u\downarrow u\uparrow d\uparrow \rangle \ +\ ( u \leftrightarrow d)\ \right].
\end{equation}

\vspace{0.2 cm}

\noindent Eq. (1) allows us to calculate the contribution of the $u$ and $d$ quarks to the proton spin:

\begin{equation}
\Delta u = \langle p \uparrow | N_{u}\uparrow \ -\ N_{u}\downarrow |p \uparrow \rangle = \frac{3}{18}(10 \ - \ 2) = \frac{4}{3}.
\end{equation}

\begin{equation}
\Delta d = \langle p \uparrow | N_{d}\uparrow \ -\ N_{d}\downarrow |p \uparrow \rangle = \frac{3}{18}(2 \ - \ 4) = -\frac{1}{3}.
\end{equation}

\vspace{0.2 cm}

\noindent where $N_{q}\uparrow$ ($N_{q}\downarrow$) represents the number of quarks of flavour $q$ with a spin projection parallel (anti-parallel) to the proton spin. Therefore, we obtain for the total helicity of quarks:


\begin{equation}
\Delta \Sigma := \Delta u + \Delta d = \left(\frac{4}{3} -\frac{1}{3}\right) = 1.
\end{equation}

\vspace{0.2 cm}

\noindent Indeed, from Eq. (4) we conclude that all the nucleon spin is carried by the valence quarks. However, in 1988 the \textbf{E}uropean \textbf{M}uon \textbf{C}ollaboration (EMC) at CERN astonished the scientific community by publishing a surprising result for the total contribution of quarks to the nucleon spin ($\Delta \Sigma$). The experimental value obtained by EMC was very small, and even compatible with zero \cite{EMC}: $\Delta \Sigma = 0.12 \pm 0.17$. This result gave rise to the so-called \textbf{spin crisis} of the nucleon because it could not be reconciled with the theoretical predictions: even applying the necessary relativistic corrections \cite{Bass}, the expectation value for $\Delta \Sigma$ is about $0.60$ \cite{Jaffe2}. Since then, polarised lepton-nucleon scattering experiments were performed at CERN by SMC \cite{SMC} and COMPASS \cite{COMPASS}, at SLAC \cite{E155}, at DESY \cite{HERMES} and at JLAB \cite{CLAS1} as well as in polarised proton-proton collisions at RICH \cite{PHENIX, STAR}. The goal is to extract the parton helicity distributions in the nucleon, using a perturbative QCD analysis. The contribution of quarks to the nucleon spin is nowadays confirmed to be $30\%$ ($\Delta \Sigma = 0.30 \pm 0.01 \pm 0.01$ \cite{COMPASS}). The fact that this number is still quite below the expected one leads us to the following question: where is the remaining part of the nucleon spin? In QCD the nucleon spin projection (in units of $\hbar$) may be decomposed into the quark and gluon helicities, $\Delta \Sigma$ and $\Delta G$, and also into their orbital angular momenta $L_{q}$ and $L_{g}$:

\begin{equation}
\frac{1}{2} =  J_{q} + J_{g} = \left(\frac{1}{2}\Delta \Sigma + L_{q}\right) + \left(\Delta G + L_{g}\right).
\end{equation}

\vspace{0.2 cm}

\noindent Knowing that the gluons were the solution to the problem of the missing momentum in the nucleon, the obvious approach to solve this \textbf{spin puzzle} (Eq. (5)) would be the determination of the gluon helicity $\Delta G$ (note that $\Delta G = \int_{0}^{1} \Delta g(x)dx$). This was a strong motivation for measuring the gluon helicity distribution as a function of the fraction $x$ of the gluon momentum, $\Delta g(x)$, in dedicated experiments like COMPASS. Another motivation is found on the fact that in QCD the U(1) anomaly generates a contribution from gluons to the measured singlet axial coupling $a_{0}(Q^2)$. This implies that $\Delta\Sigma$ is scheme dependent and may differ from the observable $a_{0} (Q^2)$, while $\Delta G$ is scheme independent at least up to the NLO approximation. In the AB (Adler-Bardeen) scheme, $\Delta\Sigma ^{AB}$ is independent of $Q^{2}$  \cite{ABscheme}. It is related to the $\overline {\textrm{MS}}$ (Minimum Subtraction) scheme through the following equation:

\begin{equation}
\Delta\Sigma^{\overline {MS}}(Q^2) = \Delta\Sigma^{AB} - n_{f}\frac{\alpha_{S}(Q^2)}{2\pi}\Delta G(Q^2).
\end{equation}

\vspace{0.2 cm}

\noindent Assuming the Ellis-Jaffe prediction of $\Delta\Sigma^{AB} \approx 0.6$, very large values of $\Delta G(Q^2)$ are required  in order to reconcile the theoretical prediction with the experimental result: $\Delta\Sigma^{\overline {MS}}(Q^2) = a_{0}(Q^{2}) \approx 0.3$.\\

\noindent Due to a limited range in $Q^2$ covered by the polarised experiments, the QCD evolution of the DGLAP equations shows a very limited sensitivity to the gluon helicity distribution, $\Delta g(x)$, and to its first moment, $\Delta G$. Therefore, the determination of $\Delta g(x)$ from QCD evolution must be complemented by direct measurements. The average gluon polarisation in a limited range of $x$, $\langle \Delta G/G \rangle _{x}$, has been determined in a model independent way, from the photon gluon fusion process, by HERMES \cite{HERMES2}, SMC \cite{SMC2} and COMPASS \cite{COMPASS4, COMPASS5}. In this work we present the latest COMPASS results obtained from an open charm analysis. Basically, the PGF process is tagged by the detection of the decay products of charmed mesons.\\

\section{The COMPASS Experiment}

\noindent COMPASS is a fixed target experiment located in the CERN North Area, at the end of the M2 beam line of the \textbf{S}uper \textbf{P}roton \textbf{S}ynchrotron (SPS) accelerator. In order to study the gluon polarisation in the nucleon, COMPASS uses a naturally polarised muon beam and a fixed polarised target. The average beam polarisation in the laboratory system is about 82\% at 160 GeV/c. This polarisation is achieved in an elegant way, resulting from parity violation decays of the mesons $\pi^{+}$  and $K^{+}$, which in turn are produced by collisions of the SPS proton beam on a thick absorber. A sketch of the M2 beam line is displayed in Fig. 1.

\begin{figure}[h]
\begin{center}
\includegraphics[totalheight=8.2cm,width=0.75\textwidth]{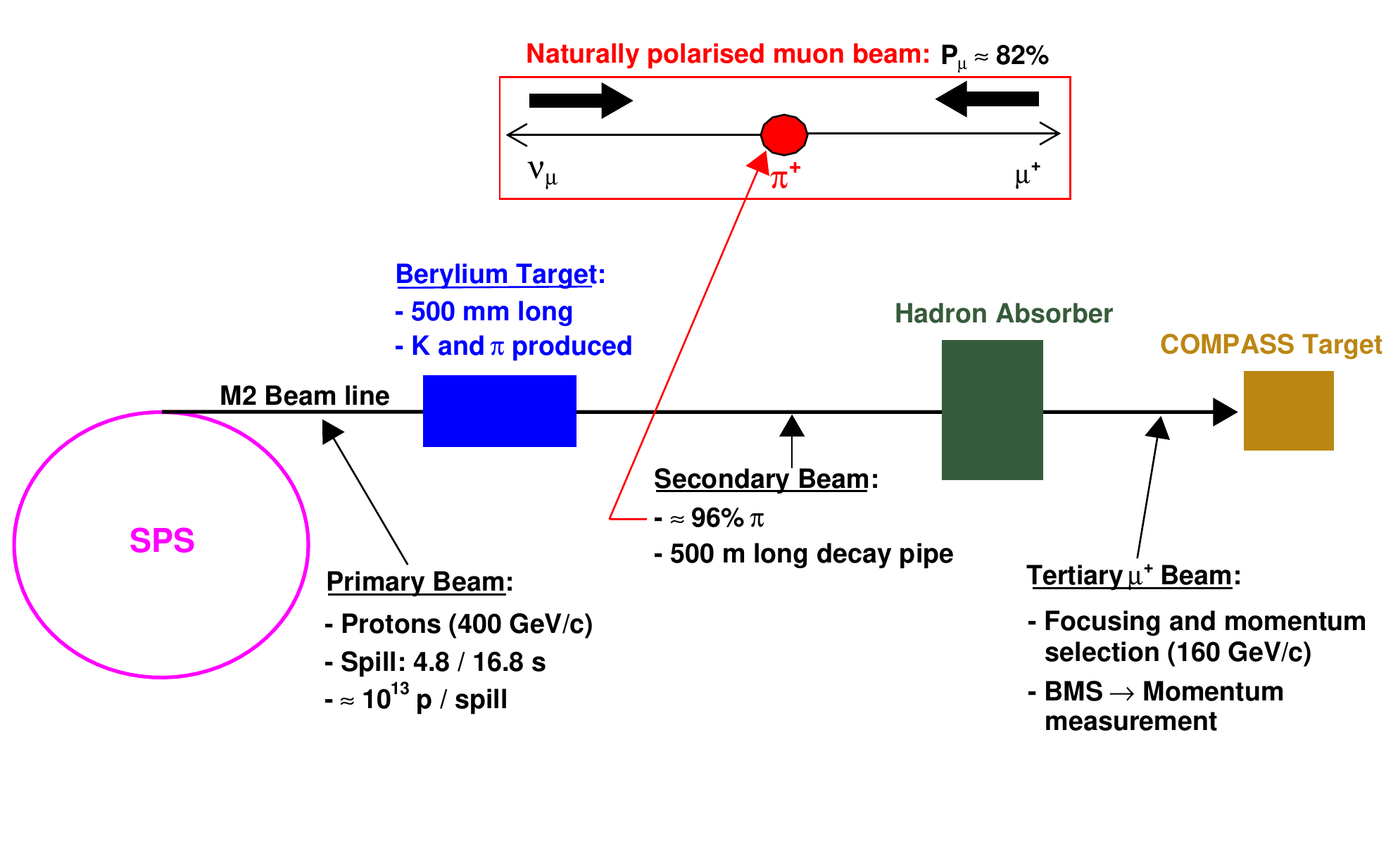}
\vspace{-1.2 cm}
\caption{Sketch of the production of a polarised muon beam for COMPASS. Note that the muons are 100\% polarised in the pion rest frame (due to helicity conservation). These polarised leptons, $\mu^{+}$, are the probes used to scan the gluon polarisation.}
\end{center}
\label{fig:Beam}       
\end{figure}

\vspace{-0.3 cm}

\noindent The COMPASS target used in 2002, 2003 and 2004 consists of two cells containing polarisable material in the solid state, each 60 cm long and 3 cm in diameter, separated by a gap of 10 cm. For the study of the gluon polarisation, the cells are longitudinally polarised in opposite directions (in order to allow for the extraction of spin asymmetries). The experimental acceptance for the resulting events is limited to an angular range of $\pm 70$ mrad. However, in the years of 2006 and 2007 the target setup was upgraded to a 3-cell system with an angular acceptance of $\pm 180$ mrad. These two improvements represent the only differences compared to the target setup used until 2004. In 2006 and 2007 the upstream and downstream cells are 30 cm long, and are both 5 cm away from a central cell of 60 cm long (their diameter is 4 cm). The former cells are longitudinally polarised in the opposite direction than that for the middle cell.\\ 
\indent Solid state targets are used, for all years of data taking, in order to achieve the high luminosity required for the physics programme of COMPASS. The target material used to study the spin structure of deuterons and protons is $^{6}$LiD (2002-2006) and NH$_{3}$ (2007), respectively. The target cells are surrounded by a powerful superconducting solenoid which is used for their polarisation. The solenoid aperture defines the acceptance for event detection in the COMPASS spectrometer. The average polarisation for the $^{6}$LiD material is $P_{t} \approx 50\%$, with a dilution factor $f \approx 0.4$. The latter accounts for the fraction of polarisable material inside each target cell. For the NH$_{3}$ material, the average polarisation is $P_{t} \approx 90\%$ and the dilution factor is  $f \approx 0.15$.\\

\noindent The particles resulting from inelastic scattering collisions, i.e., the 160 GeV/c muon beam scattered off the $^{6}$LiD or NH$_{3}$ material of the COMPASS target, are then detected in a two stage spectrometer whose detailed description can be found in Ref. \cite{Spectro}. Basically, each spectrometer is defined by a large magnet surrounded by several detectors (tracking, calorimeters, etc.). The first one is the large angle spectrometer, dedicated to the reconstruction of low momentum tracks, whereas the second spectrometer is designed to detect particles at smaller angles. A sketch of the COMPASS spectrometer is displayed in Fig. 2.

\vspace{0.2 cm}

\begin{figure}[h]
\begin{center}
\includegraphics[totalheight=9.0cm,width=0.7\textwidth]{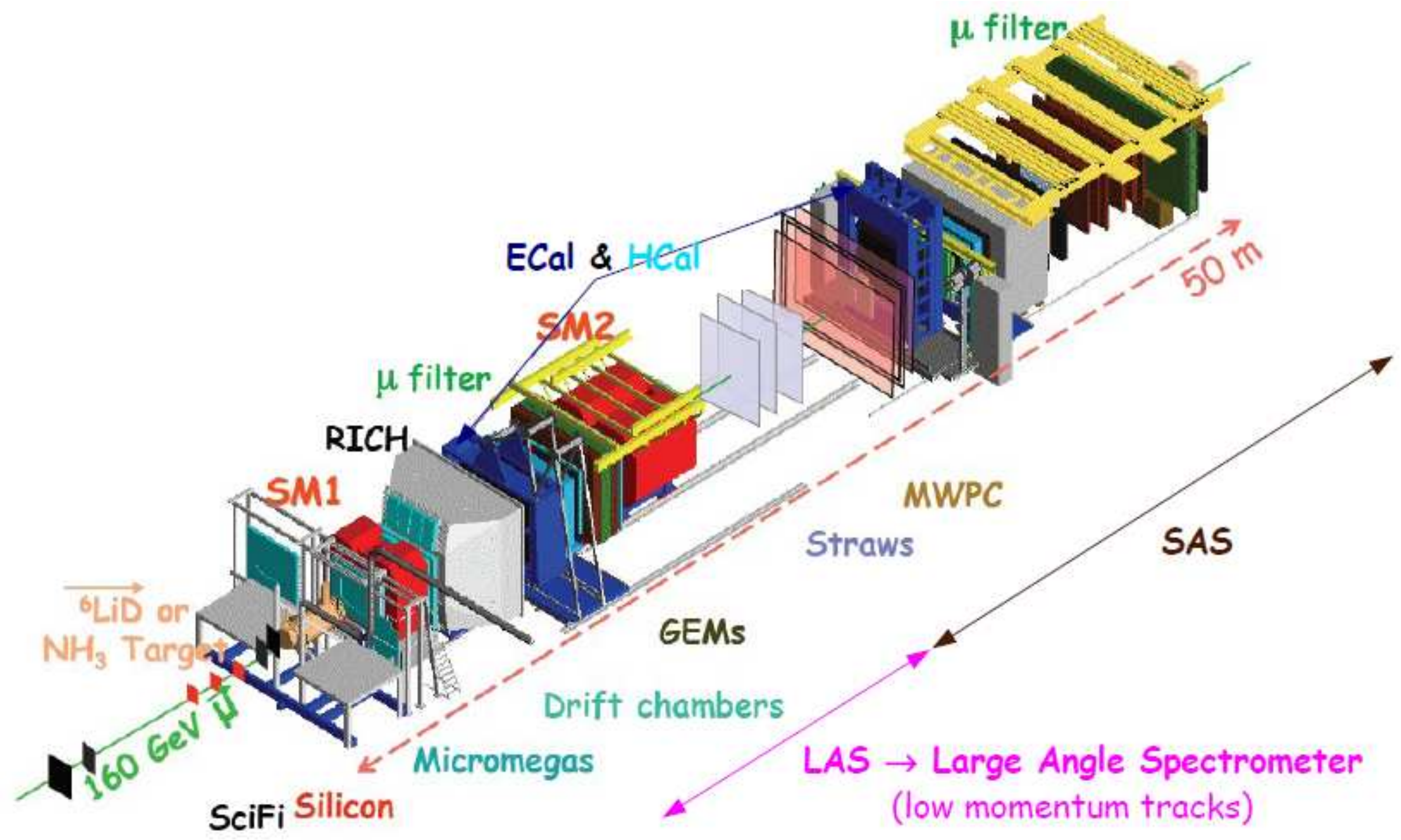}
\vspace{0.2 cm}
\caption{Artistic view of the COMPASS spectrometer during the 2004 data taking. The polarised muon beam enters from the left, and after being tracked by a set of SciFi and Silicon detectors collides in the polarised target. The products of the reaction are detected by a two stage spectrometer, using the SM1 and SM2 magnets.}
\end{center}
\label{fig:2}       
\end{figure}

\vspace{0.2 cm}

\noindent The most common charged particles ($e$, $\pi$, $k$ and $p$) can be identified by a Ring Imaging CHerenkov (RICH) detector, which is located in the first spectrometer. It consists of several multi-wire proportional chambers, containing CsI photocathodes, which detect the UV components of the Cherenkov light. In 2006, there was a considerable upgrade of the RICH: the central part of the detector was replaced by multi-anode photomultiplier tubes, yielding a larger number of photons detected (by extending the range of detection to the visible wavelength) together with a much faster response (reducing the uncorrelated background of halo muons). Together with a readout electronics refurbishment, this upgrade allowed for a much cleaner particle identification.

\section{Event selection}

\noindent The cleanest way to measure the gluon helicity in the nucleon is to analyse the events from polarised inelastic scattering experiments. In order to be sensitive to $\Delta g(x)$, one must tag a process involving a polarised lepton-gluon interaction.  In COMPASS, one of the possibilities to do it is to reconstruct charmed mesons:

\begin{figure}[!h]
\begin{center}
\includegraphics[totalheight=7.2cm,width=0.55\textwidth]{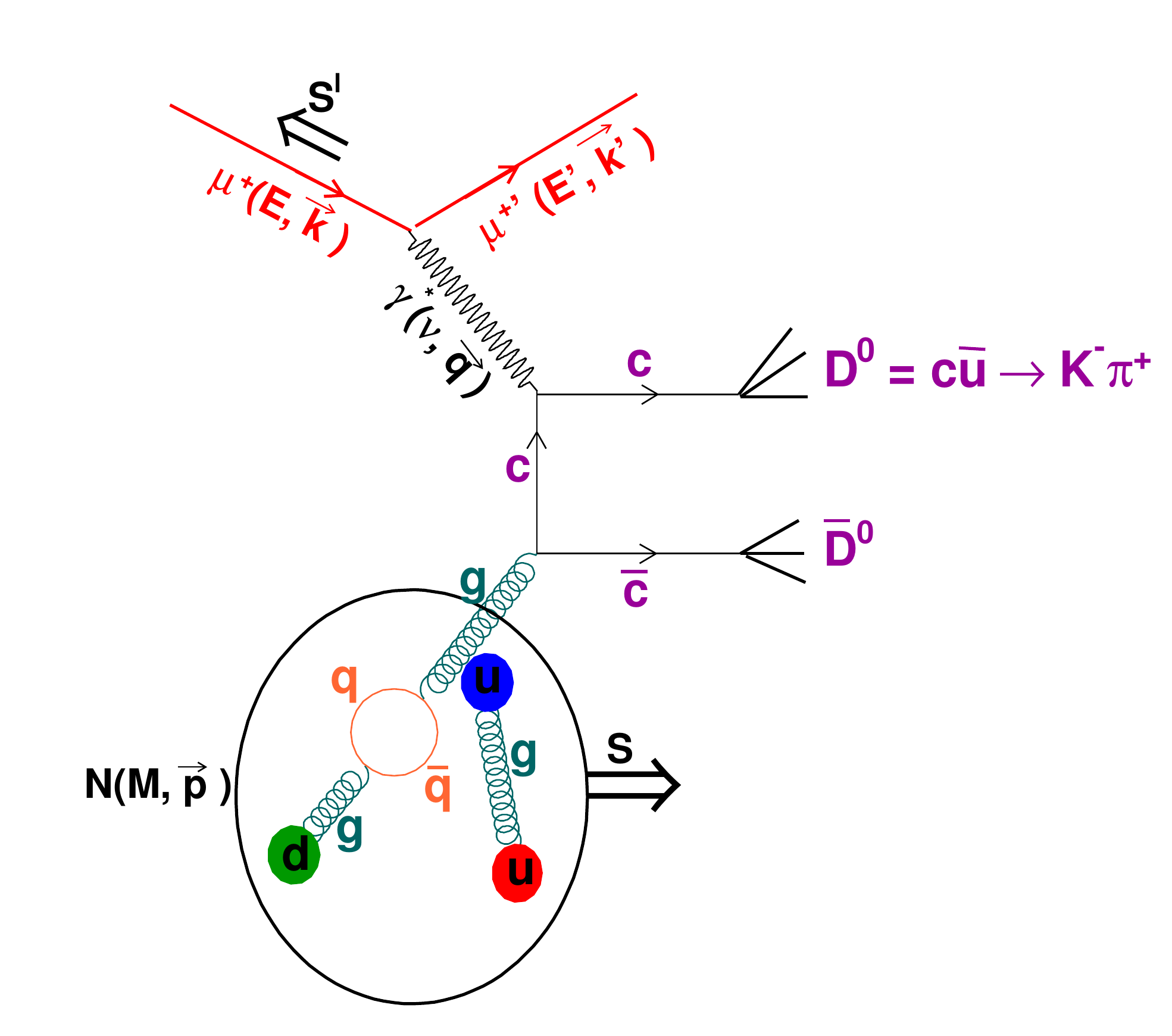}
\end{center}
\label{fig:PGF1}
\caption{Open charm production from a polarised PGF process (at LO)}
\end{figure}

\vspace{0.3 cm}

\noindent This method provides a clean signature of PGF events, because the intrinsic charm content of the nucleon is negligible in the kinematic domain of COMPASS (cf. Fig. 4). In this analysis the PGF process is tagged using the $D^{0}$ meson production. This production is limited to a range of small $x_{Bj}$, $x_{Bj} < 0.1$, because at COMPASS kinematics the cross-section for PGF events decreases rapidly with $x_{Bj}$. \\

\begin{figure}[!h]
\begin{center}
\includegraphics[totalheight=6.8cm,width=0.51\textwidth]{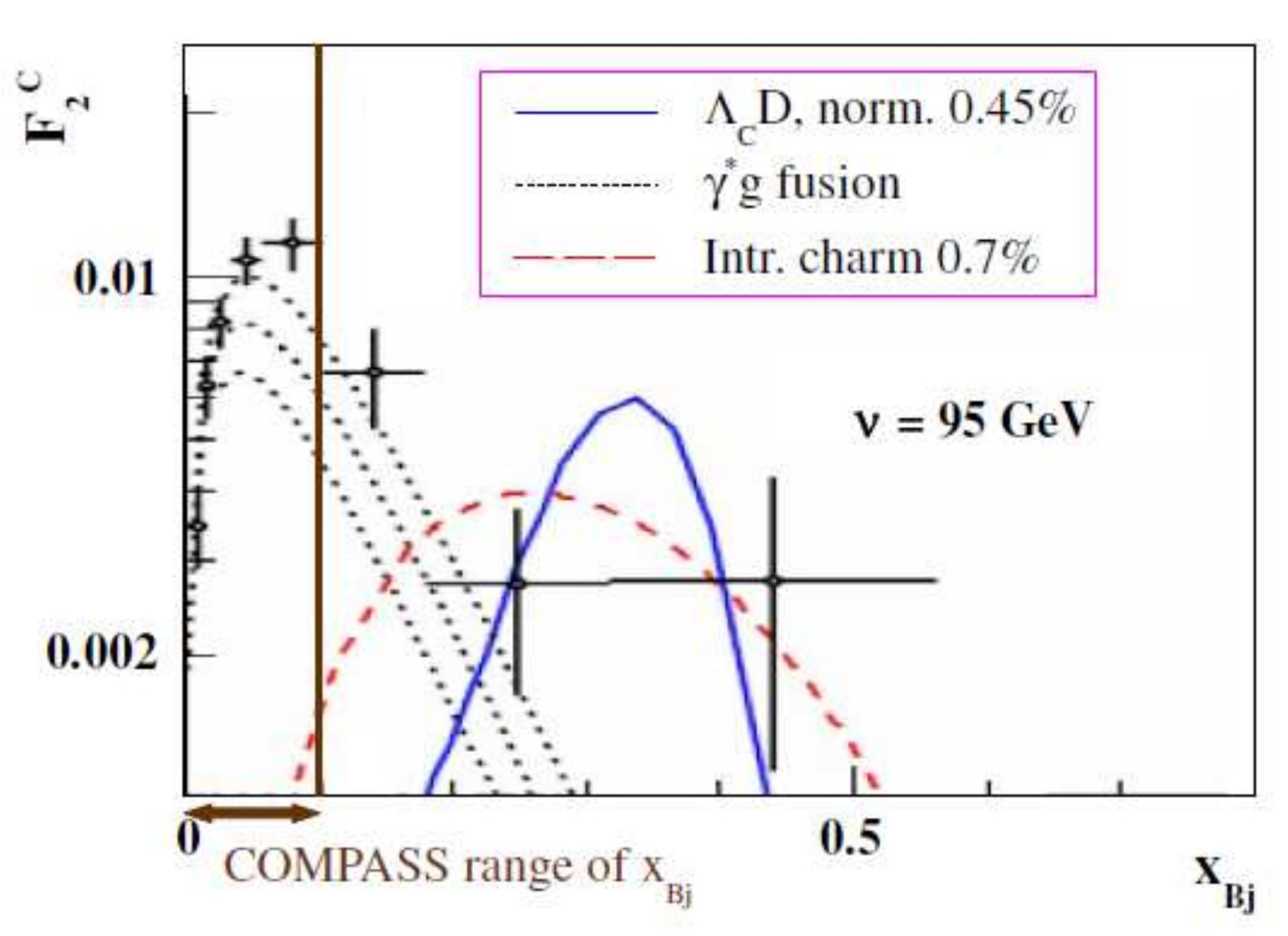}
\end{center}
\label{fig:IC}
\vspace{-0.55 cm}
\caption{Intrinsic charm predictions obtained for the virtual photon energy $\nu$ equal to its average value for the COMPASS data: $\langle \nu \rangle = 0.95$ GeV. The experimental points represent measurements of the charm structure function by the collaboration EMC \cite{EMC1}, and the black dotted curves are theoretical fits for the PGF process using different renormalisation and factorisation schemes \cite{PGFfits}. The blue curve shows the intrinsic charm prediction assuming a nucleon fluctuation into charmed meson-hadron pairs \cite{Meson-Hadron}. The red dashed curve shows the intrinsic charm prediction assuming charm fluctuations at the partonic level \cite{Brodsky}.}
\end{figure}

\vspace{0.2 cm}

\noindent The COMPASS spectrometer is designed to reconstruct the $D^{0}$ mesons through the invariant mass of their decay products, $K\pi$ pairs, and for that purpose the RICH detector plays an extremely important role: the requirement of proper identification of kaon and pion candidates reduces significantly the combinatorial background underlying the resonance, which is centred on the $D^{0}$ mass, cf. Fig. 5. In addition, the following kinematic cuts are applied: on the fraction of the virtual photon energy carried by the $D^{0}$, $0.2 < z_{D^{0}} < 0.85$, and on the angle between the charmed meson direction and the resulting kaon in the $D^{0}$ centre-of-mass system, $|\textrm{cos}\, \theta^{*}| < 0.65$. They are important to reduce the contamination of the PGF sample by events coming from processes that involve the fragmentation of a struck quark, because most of these events are almost collinear with the virtual photon direction or have a $z_{D^{0}}$ values close to zero. The combinatorial background can be further reduced using the following channel: $D^{*} \rightarrow D^{0}\pi_{slow}$ with $D^{0} \rightarrow K\pi$ ($D^{0}$ tagged with a $D^{*}$). By applying a cut on the difference of the reconstructed masses for the $D^{*}$ and $D^{0}$ mesons, 3.2  MeV/c$^2 < M_{D^*}^{rec} - M_{D^0}^{rec} - M_{\pi} <$ 8.9  MeV/c$^2$, one can verify that there is not much room left for the slow pion momentum.  Since the above mass difference is reconstructed accurately\footnote[2]{Note that the cryogenic solid state target of COMPASS does not permit the distinction between the $D^{*}$ and $D^{0}$ decay vertices (or between the primary and the decay vertices of a $D^{0}$).}, it can be used as a cut to significantly improve the purity of the $D^{0}$ signal. Consequently, three new channels of lower purity can also be studied (all tagged with a $D^{*}$): $D^{0} \rightarrow K\pi\pi^{0}$, $D^{0} \rightarrow K_{sub}\pi$, and $D^{0} \rightarrow K\pi\pi\pi$. The resonance observed in the first channel emerges from the combinatorial background as a 'bump', centred around -250 MeV/c$^{2}$ in the $D^{0} \rightarrow K\pi$ spectra, due to the fact that the extra $\pi^{0}$ is not directly reconstructed for this analysis. The second channel represents the $D^{0}$ candidates without RICH response for the kaon mass hypothesis (sub-threshold kaons with p(K) $<$ 9 GeV/c), and the last one involves another $D^{0}$ decay mode which helps to improve the statistical precision of the $\Delta g/g$ measurement. The final $D^{0}$ samples used in this analysis are shown in Fig. 5.




\section{Method to extract the gluon polarisation}

\noindent The number of $D^{0}$ candidates collected in a given target cell and time interval is:

\begin{equation}
\frac {d^{k}n}{dmd\xi} = a\phi \eta(s + b)\left[1 + P_tP_{\mu}f\left(\frac {s}{s+b}A^{\mu N \rightarrow \mu 'D^{0}X} + \frac {b}{s+b}A_B\right)\right]. 
\end{equation} 

\vspace{0.1 cm}

\noindent $A^{\mu N \rightarrow \mu 'D^{0}X}$ is the longitudinal double spin asymmetry of the differential cross-section for events with a $D^{0}$ or $\bar{D}^{0}$ in the final state, and $A_{B}$ is the corresponding asymmetry originating from combinatorial background events. Furthermore, $m = M_{K\pi}$ (or $m = M_{K\pi\pi\pi}$) and the symbol $\xi$ denotes a set of $k - 1$ kinematic variables describing an event ($p_{T}^{D^{0}}$, $E_{D^{0}}$,  ...), whereas $a$, $\phi$ and $\eta$ are the spectrometer acceptance, the incident muon flux integrated over the time interval, and the number of target nucleons respectively. The differential unpolarised cross-sections for signal and background events folded with the experimental resolution as a function of $m$ and $\xi$ are represented by $s = s(m, \xi)$ and $b = b(m, \xi)$ respectively. The ratio $s/(s+b)$ represents the signal purity of the $D^{0}$ mass spectra. The information on the gluon polarisation is contained in the muon-nucleon asymmetry $A^{\mu N}$. The latter is defined in LO QCD as a convolution between the ratio of polarised/unpolarised partonic cross-sections, $(\Delta \hat{\sigma}/\hat{\sigma})^{\mu g \rightarrow \mu'c\bar{c}}$, and the ratio of polarised/unpolarised gluon structure functions ($\Delta g/g$):

\begin{equation}
A^{\mu N} = \langle \hat{a}_{LL} \rangle\frac{\Delta g}{g} \ \ \ \ \ \textrm{with} \ \ \ \ \ \hat{a}_{LL} \equiv \frac{\Delta \hat{\sigma}_{\mu g}}{\hat{\sigma}_{\mu g}}  = \left(\frac{\hat{\sigma}^{\overleftarrow{\Rightarrow}}_{\mu g} - \hat{\sigma}^{\overleftarrow{\Leftarrow}}_{\mu g}}{\hat{\sigma}^{\overleftarrow{\Rightarrow}}_{\mu g} + \hat{\sigma}^{\overleftarrow{\Leftarrow}}_{\mu g}}\right).
\end{equation}

\noindent  Four equations like Eq. (7) are defined, i.e. one equation for each cell and spin configuration of the target. Since the factors $s/(s+b)$ and $a_{LL} (\equiv \langle \hat{a}_{LL} \rangle)$ have a large dispersion, a weighting method is used to minimise the statistical error. The signal weight, $\omega_{S} = P_{\mu}fa_{LL}[s/(s+b)]$, and the background weight, $\omega_{B} = P_{\mu}fD[b/(s+b)]$,

\begin{figure}[|h]
\begin{center}$
\begin{array}{cc}
\includegraphics[totalheight=5.8cm,width=0.45\textwidth]{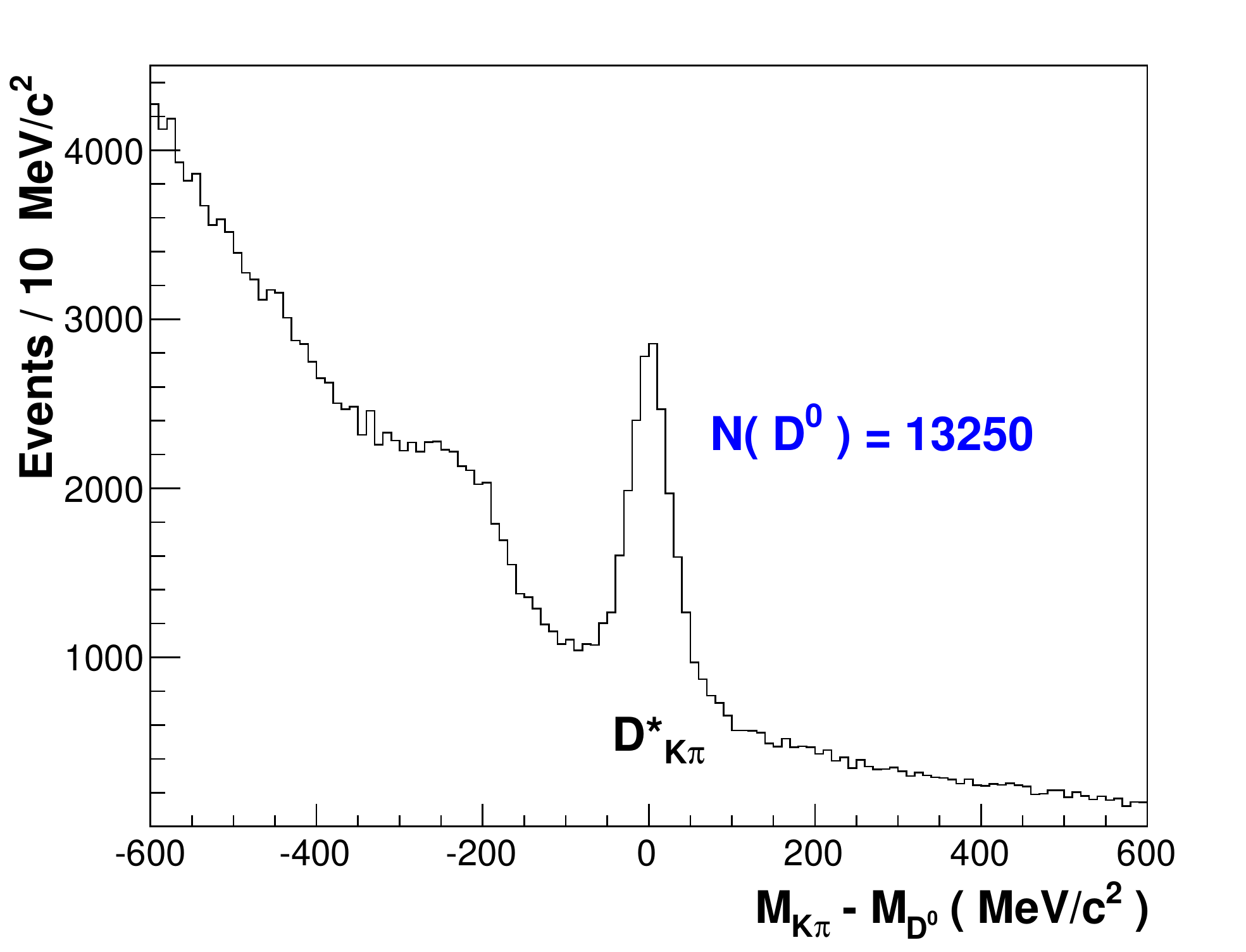} &
\includegraphics[totalheight=5.8cm,width=0.45\textwidth]{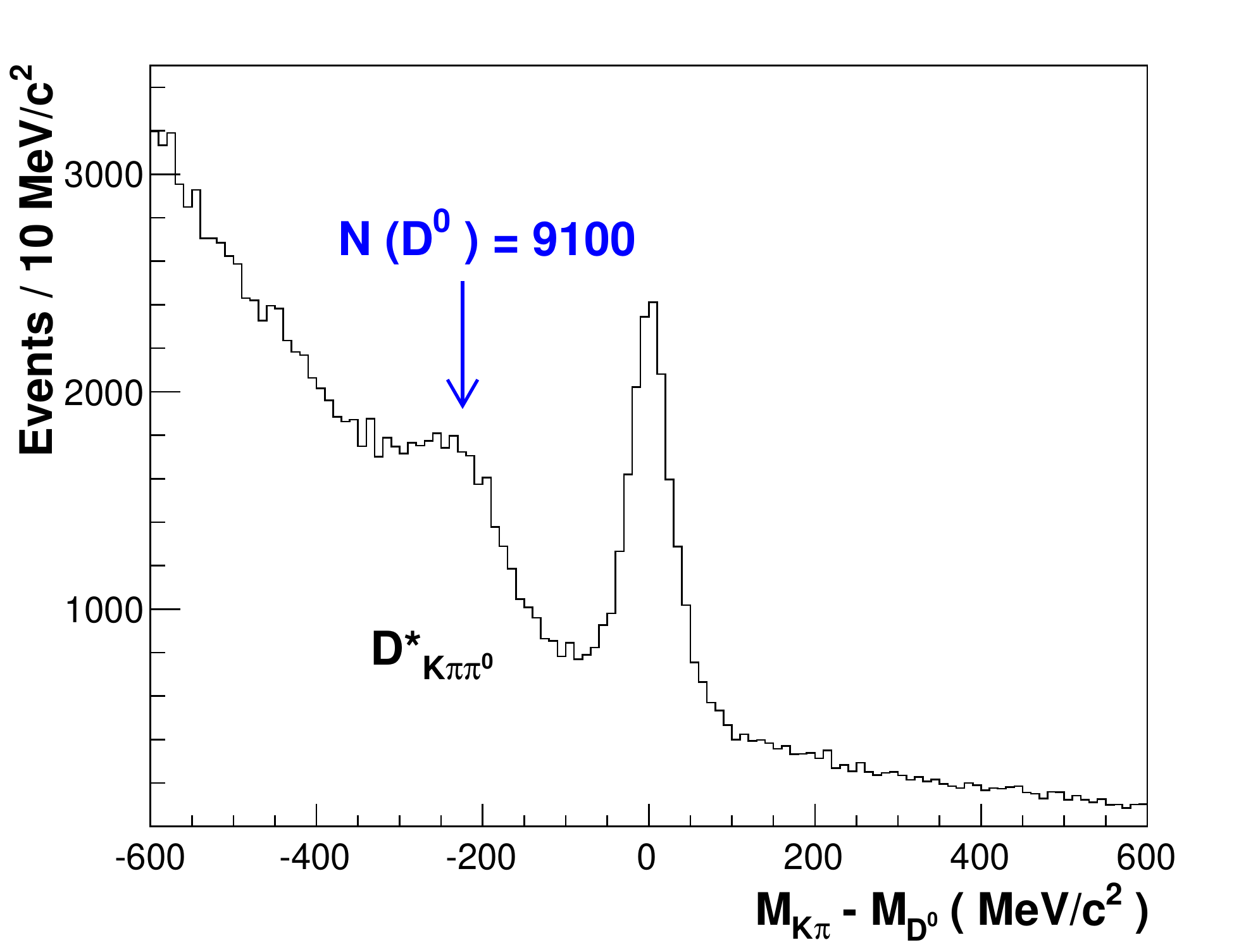} \\
\includegraphics[totalheight=5.8cm,width=0.45\textwidth]{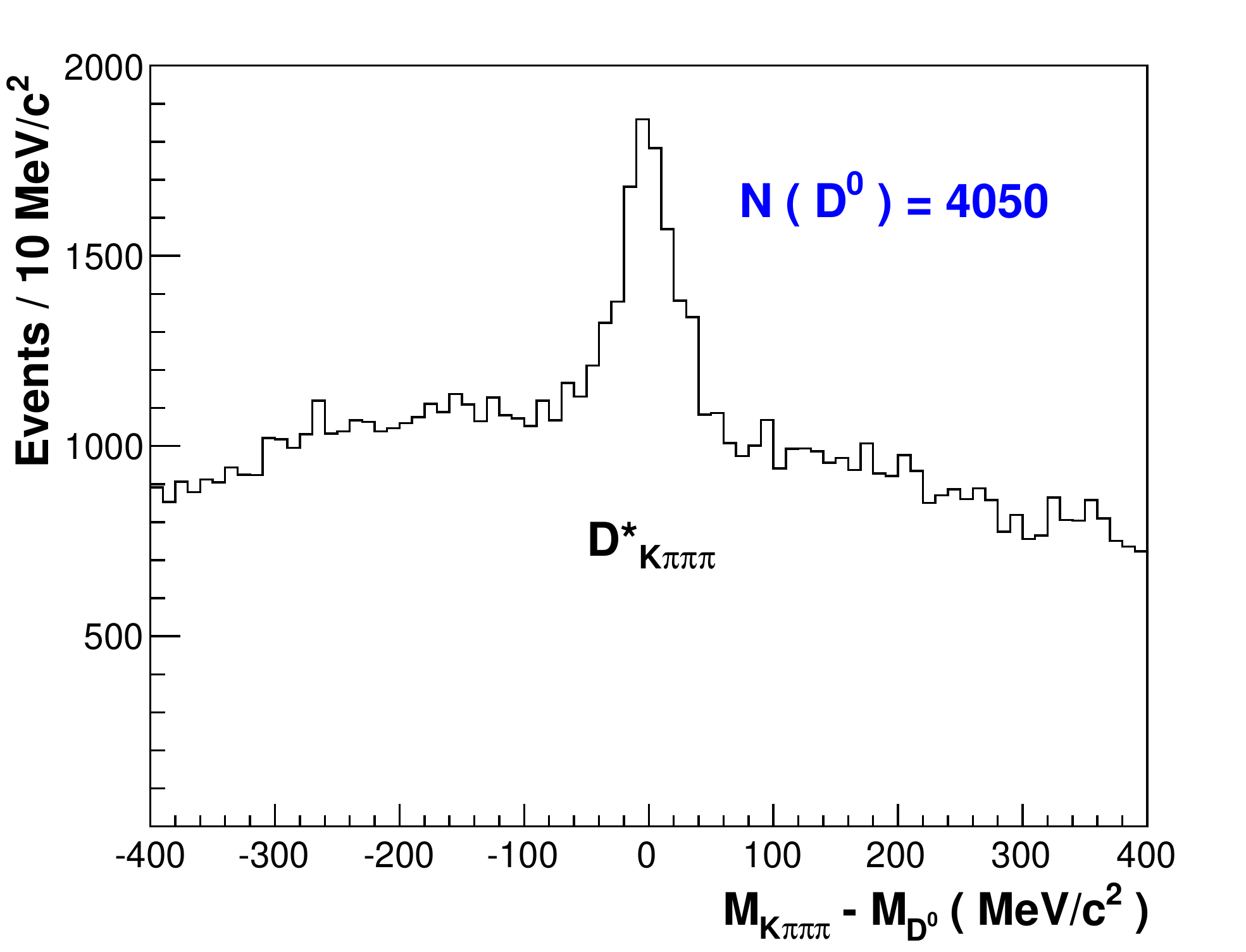} &
\includegraphics[totalheight=5.8cm,width=0.45\textwidth]{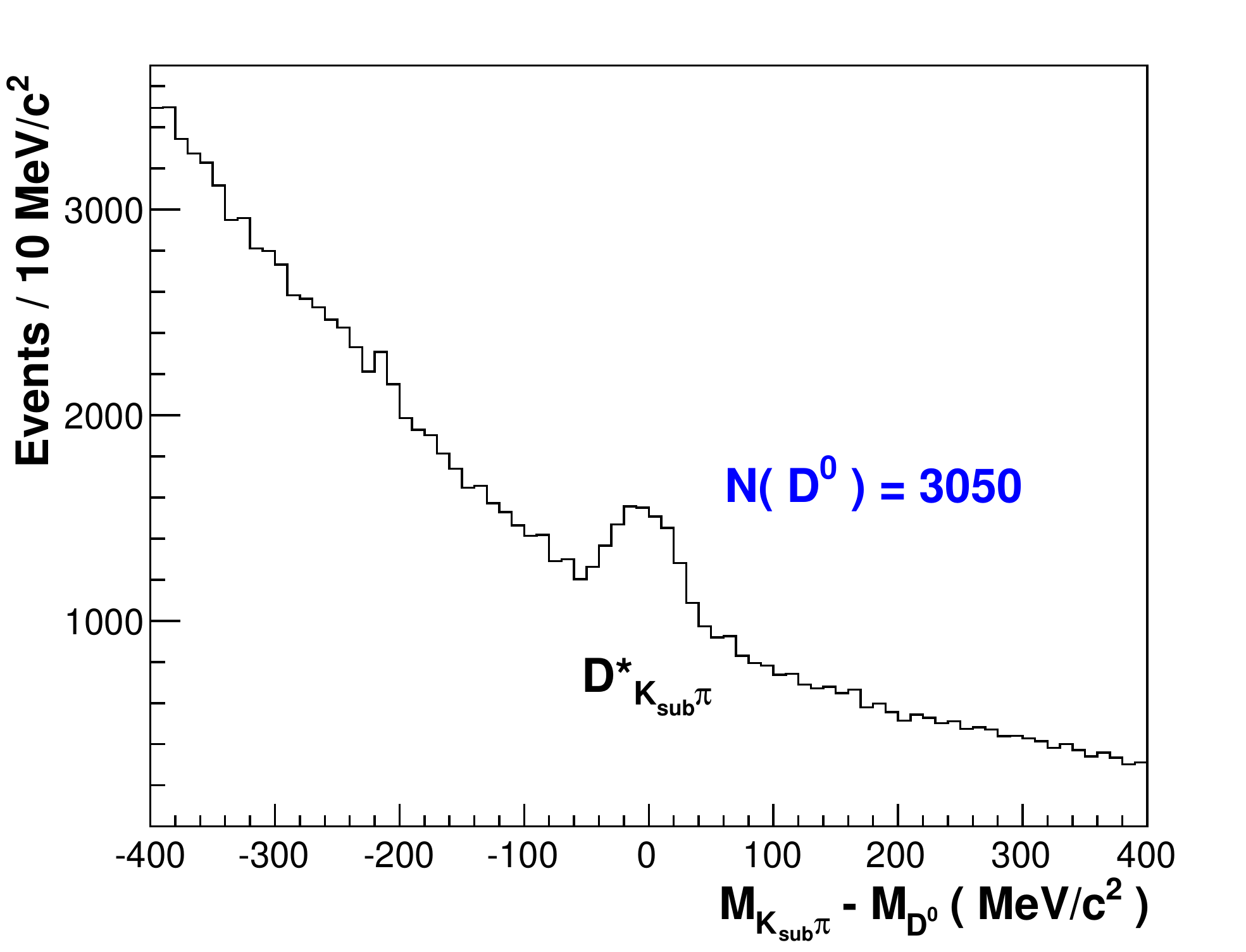} \\
\includegraphics[totalheight=5.8cm,width=0.45\textwidth]{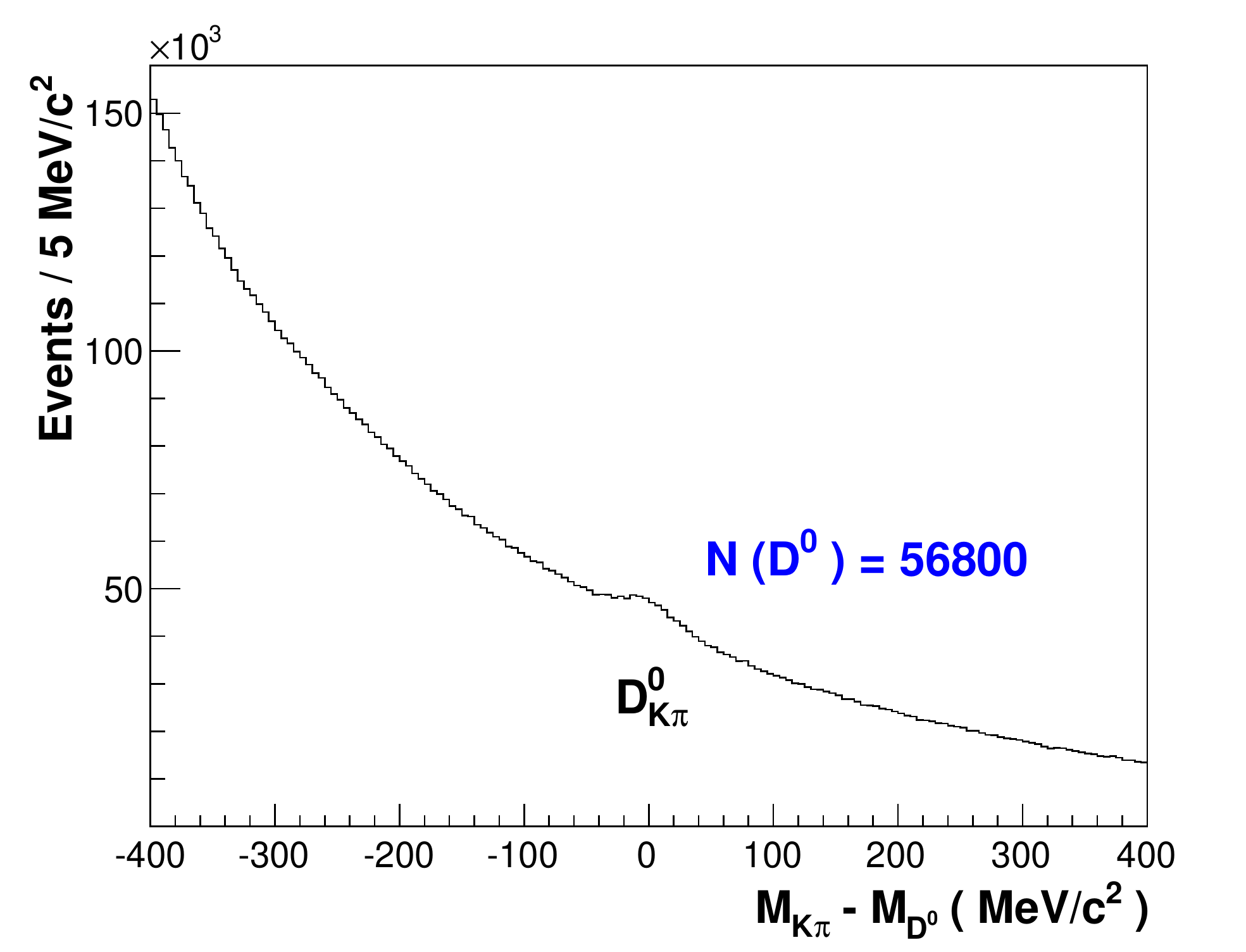} \\
\end{array}$
\end{center}
\caption{Invariant mass spectra for the $D^{*}_{K\pi}$, $D^{*}_{K\pi\pi^{0}}$, $D^{*}_{K\pi\pi\pi}$, $D^{*}_{K_{sub}\pi}$ and $D^{0}_{K\pi}$ samples. The symbol $D^{*}$ denotes a $D^{0}$ meson tagged with a $D^{*}$. The subscript notation indicates the final state of a $D^{0}$ decay. Both deuteron and proton data are included in these samples. }
\label{fig:D0_mass}
\end{figure}

\clearpage

\noindent are used, where $D$ represents the transfer of polarisation from the muon to the virtual photon. By weighting the events from each of four equations (like Eq. (7)) with $\omega_S$, and then similarly with $\omega_B$, it becomes possible to solve the system of 8 equations in order to extract $\Delta g/g$ \footnote[3]{COMPASS extracts $\Delta g/g$ instead of $\Delta g$ because experimentally it is much simpler to measure asymmetries than cross-section differences.}. Only 7 unknowns exist in this system of 8 equations: $\Delta g/g$, $A_B$ and 5 independent quantities involving the product $a\phi \eta$ (see Ref. \cite{Celso}). However, to solve the system in an optimal way one needs $a_{LL}$ and $s/(s+b)$ for every event.\\

\noindent In the next-to-leading order (NLO) approximation, the physical asymmetry is decomposed as follows:

\begin{equation}
A^{\mu N} = a_{LL}^{PGF}\frac{\Delta g}{g} +  a_{LL}^{q}A_{1},
\end{equation}

\noindent where in $a_{LL}^{PGF}$ all the soft, virtual and gluon bremsstrahlung corrections to the LO-PGF process are included (cf. diagrams a), b) and c) of Fig. 6). The asymmetry $a_{LL}^{q}$ accounts for the presence of physical background inside the $D^{0}$ resonance. It corresponds to an interaction between the virtual photon and a light quark, which subsequently (or immediately before) emits a gluon that produces a $c\bar{c}$ pair (cf. diagram d) of Fig. 6) . One of these charmed quarks may fragment into a non-PGF $D^{0}$. Finally, $A_{1}$ represents the inclusive asymmetry which is experimentally well known \cite{Celso1}. The contamination described by the second term of Eq. (9) is negligible and it appears only at NLO or higher orders of approximation in QCD.\\
\indent Using Eqs. (7) and (8) or Eqs. (7) and (9) one is able to determine $\Delta g/g$ at the LO or NLO approximations. In order to make possible the use of COMPASS results in global analyses, a set of virtual photon asymmetries, $A^{\gamma^{*} N \rightarrow \mu 'D^{0}X} = A^{\mu N}/D$, was determined in bins of $p_{T}^{D^{0}}$ and $E_{D^{0}}$. They are obtained from Eq. (7) using as a signal weight the factor $\omega^{'}_{S} = P_{\mu}fD[s/(s+b)]$. Note that determined values of $A^{\gamma^{*} N}$ are independent of $a_{LL}$ and, therefore, their determination does not depend on a theoretical interpretation. The criterion for the choice of the binning is to provide asymmetries which are independent of the COMPASS acceptance for the PGF process. A table containing values of $A^{\gamma^{*} N}(p_{T}^{D^{0}}, E_{D^{0}})$ can be found in Ref. \cite{Celso}.\\

\begin{figure}[!h]
\begin{center}
\includegraphics[totalheight=6.0cm,width=0.43\textwidth]{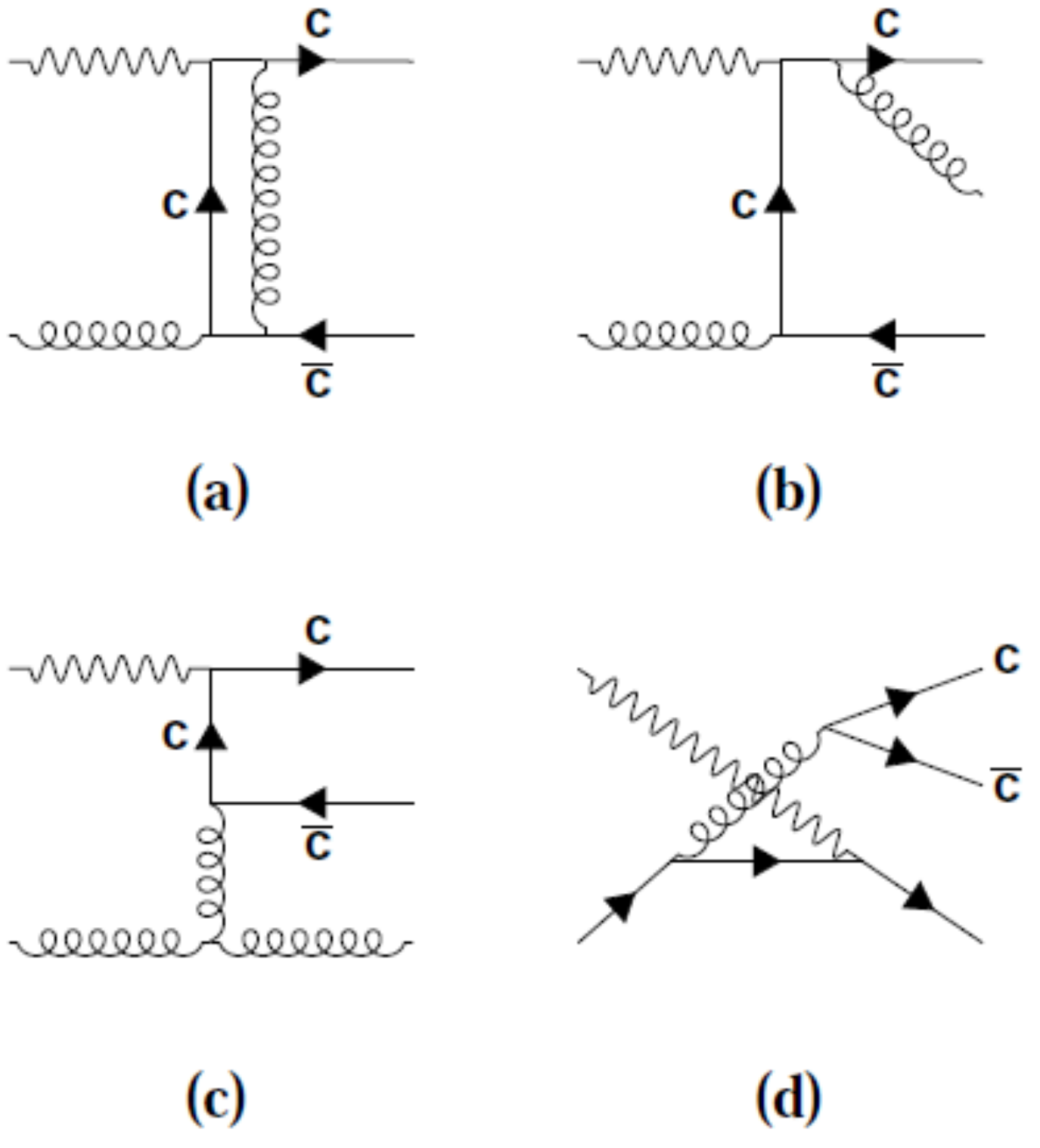}
\end{center}
\label{fig:NLO}
\vspace{-0.3 cm}
\caption{Examples of the NLO processes contributing to the muoproduction of the $c\bar{c}$ pair: a) virtual correction, b), c) gluon bremsstrahlung, d) light quark background. The remaining diagrams can be found in Ref. \cite{Celso_Thesis}.}
\end{figure}

\vspace{0.2 cm}

\section{Analysing power}

\noindent The analysing power for the open charm production, $a_{LL}$, is dependent on the full knowledge of partonic kinematics. Consequently, this asymmetry is not experimentally accessible because only one $D^{0}$ meson is reconstructed per event; the information associated with the second charm quark is lost. Nevertheless, $a_{LL}$ can be obtained from a Monte Carlo generator for production of heavy flavours (AROMA). 

\subsection{The analysing power at LO in QCD}

\noindent The AROMA generator is used without parton showers in order to generate $D^{0}$ events from LO-PGF processes. After a full Monte Carlo chain, where the generated events are constrained to the COMPASS acceptance, all $D^{0}$ mesons are reconstructed in the simulated spectrometer. This model reproduces very well the measured $D^{0}$ kinematics at COMPASS \cite{Charm2012}. The polarised and unpolarised partonic cross-sections are calculated using the information from the generator, and then the kinematic dependences of $a_{LL}$ are parametrised with the help of a Neural Network \cite{NNet}. Finally, the Monte Carlo parametrisation is used to obtain $a_{LL}$ for each real data event.

\subsection{The analysing power at NLO in QCD}

\noindent The phase space needed for NLO real gluon emission processes, $\gamma^{*} g \rightarrow c\bar{c}g$, is simulated through parton showers included in the AROMA generator. For every simulated event, an integration is performed over the energy of the unobserved gluon in the NLO emission process. The details about the upper limit of integration can be found in Ref. \cite{Celso}. This integration reduces the differential cross-section for a three-body final state ($c\bar{c}g$) to a two-body one ($c\bar{c}$), which can be combined with the LO cross-section ($c\bar{c}$, PGF) and the two-body virtual and soft NLO corrections. The procedure guarantees a correct infra-red divergence cancellation. In this way, the total partonic cross-section at NLO is calculated on an event-by-event basis for the spin averaged as well as spin dependent case, and consequently $a_{LL}^{PGF}$ at NLO accuracy is obtained. The same procedure is applied for the correction originating from $a_{LL}^{q}$. In the NLO approximation, $\Delta g/g$ is estimated from $A^{\gamma^{*} N}$ using $\langle a_{LL}^{PGF} \rangle$ and $\langle a_{LL}^{q} \rangle$. In Fig. 7 one can see that the NLO corrections are important for this open charm analysis.

 
\begin{figure}[!h]
\begin{center}
\includegraphics[totalheight=7.0cm,width=0.5\textwidth]{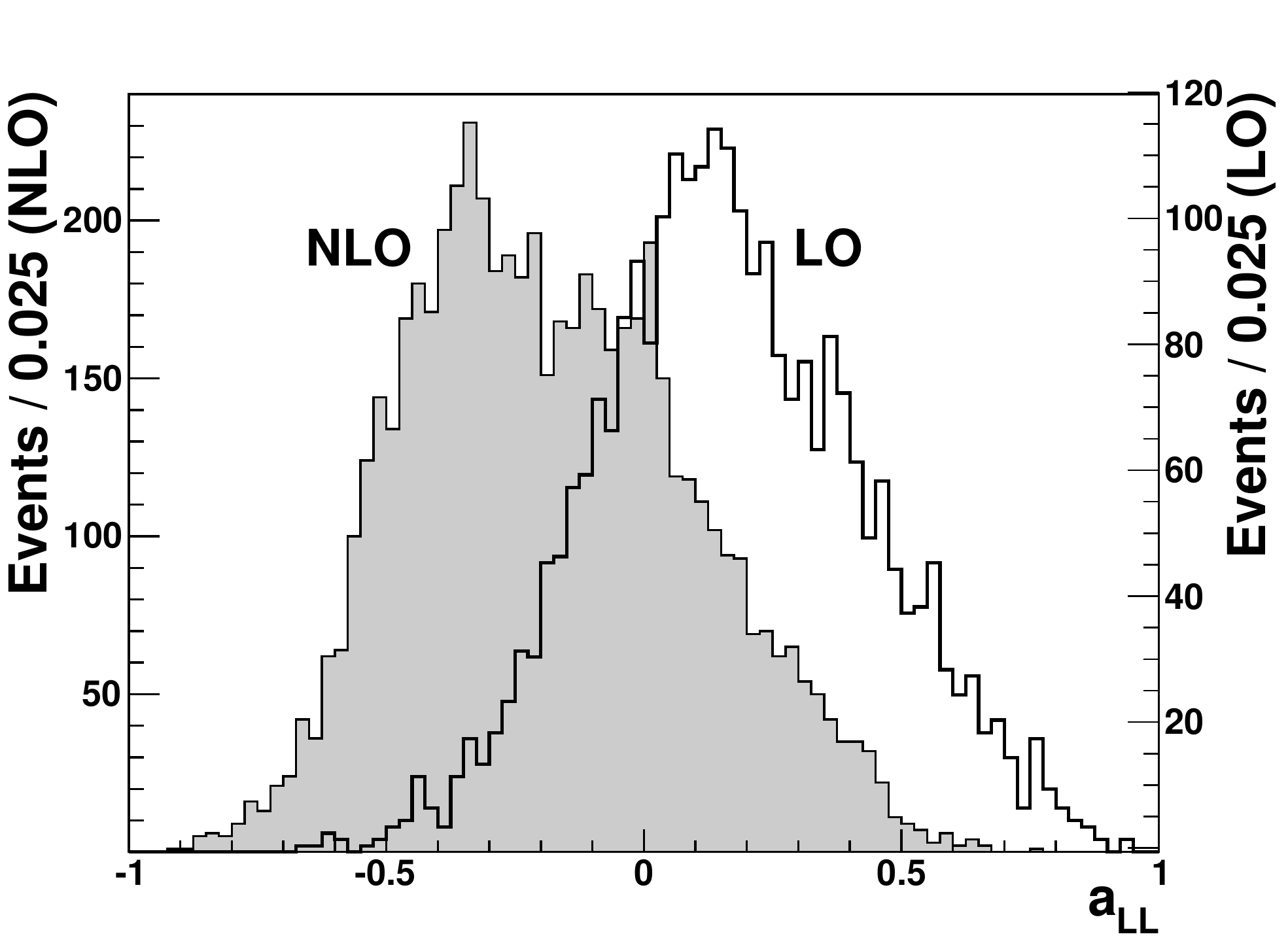}
\end{center}
\label{fig:aLL1}
\vspace{-0.3 cm}
\caption{Distribution of the analysing power $a_{LL}$ at LO and at NLO approximations for the simulated $D^{*}_{K\pi}$ events.}
\end{figure}

\section{The signal purity}

\noindent The Neural Network (NN) described in Ref. \cite{NNet} is also used to parametrise $s/(s+b)$ on real data. Here, the goal is to obtain $D^{0}$ probabilities for every event. In Fig. 8 one can see the outcome of this parametrisation: the $D^{0}$ mass spectrum reveals a probability behaviour in bins of $[s/(s+b)]_{\textrm{NN}}$, i.e. its purity increases towards $[s/(s+b)]_{\textrm{NN}}$ = 1. Consequently, the statistical precision of $\Delta g/g$ is significantly improved due to a good separation of the physical events from the combinatorial background. To accomplish this, two data sets are used as inputs to the Network. The first one contains the $D^{0}$ signal and the combinatorial background events. These events are called good charge combination ones ($gcc$), referring to the charges of particles from $D^{0}$ decays, and they are selected as described in Sec. 3. The second set, the wrong charge combination events ($wcc$), is selected in a similar way except that the sum of charges of corresponding particles should be different from zero. It contains\\

\begin{figure}[!h]
\begin{center}
\includegraphics[totalheight=15.0cm,width=0.9\textwidth]{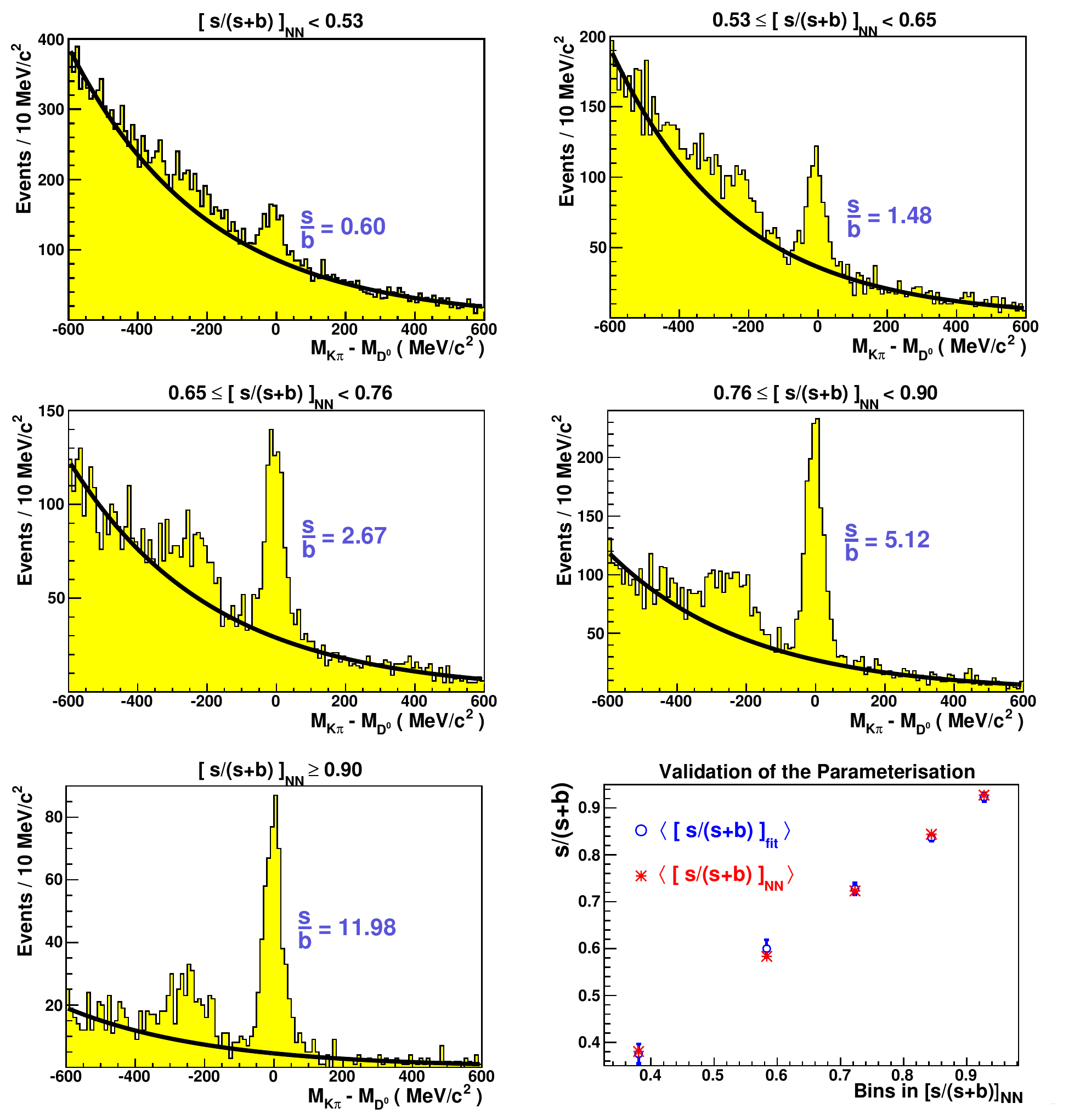}
\end{center}
\label{fig:NNet}
\vspace{-0.3 cm}
\caption{The $K\pi$ invariant mass spectra in bins of the NN signal purity, $[s/(s+b)]_{\textrm{\small{NN}}}$, for the $D^{*}_{K\pi}$ sample (example from 2006 data). The last panel shows a comparison of the two purities, $\langle [s/(s+b)]_{\textrm{NN}} \rangle$ and $[s/(s+b)]_{\textrm{fit}}$. The latter is obtained from a fit to the mass spectra. Curves show the background component of the invariant mass fits. The significance of the $D^{*}_{K\pi}$ signal is shown as the ratio $s/b$. }
\end{figure}

\vspace{0.2 cm}

\noindent only background events and is used as a background model. The NN performs a multi-dimensional comparison of $gcc$ and $wcc$ events in a $\pm 40$ MeV/c$^{2}$  mass window around the $D^{0}$ mass. Within the $gcc$ set, signal events are distinguished from combinatorial background by exploiting differences between the $gcc$ and $wcc$ sets in the shapes of distributions of kinematic variables as well as multi-dimensional correlations between them. An example of a properly chosen variable for the Network is the kaon angular distribution in the $D^{0}$ centre-of-mass system, as shown in Fig. 9. The distributions in the side band bins, shown in the right plot of Fig. 9, illustrate the good quality of the background model.

\begin{figure}[!h]
\begin{center}
\begin{minipage}[c]{0.46\linewidth}
\includegraphics[totalheight=6.3cm,width=0.96\textwidth]{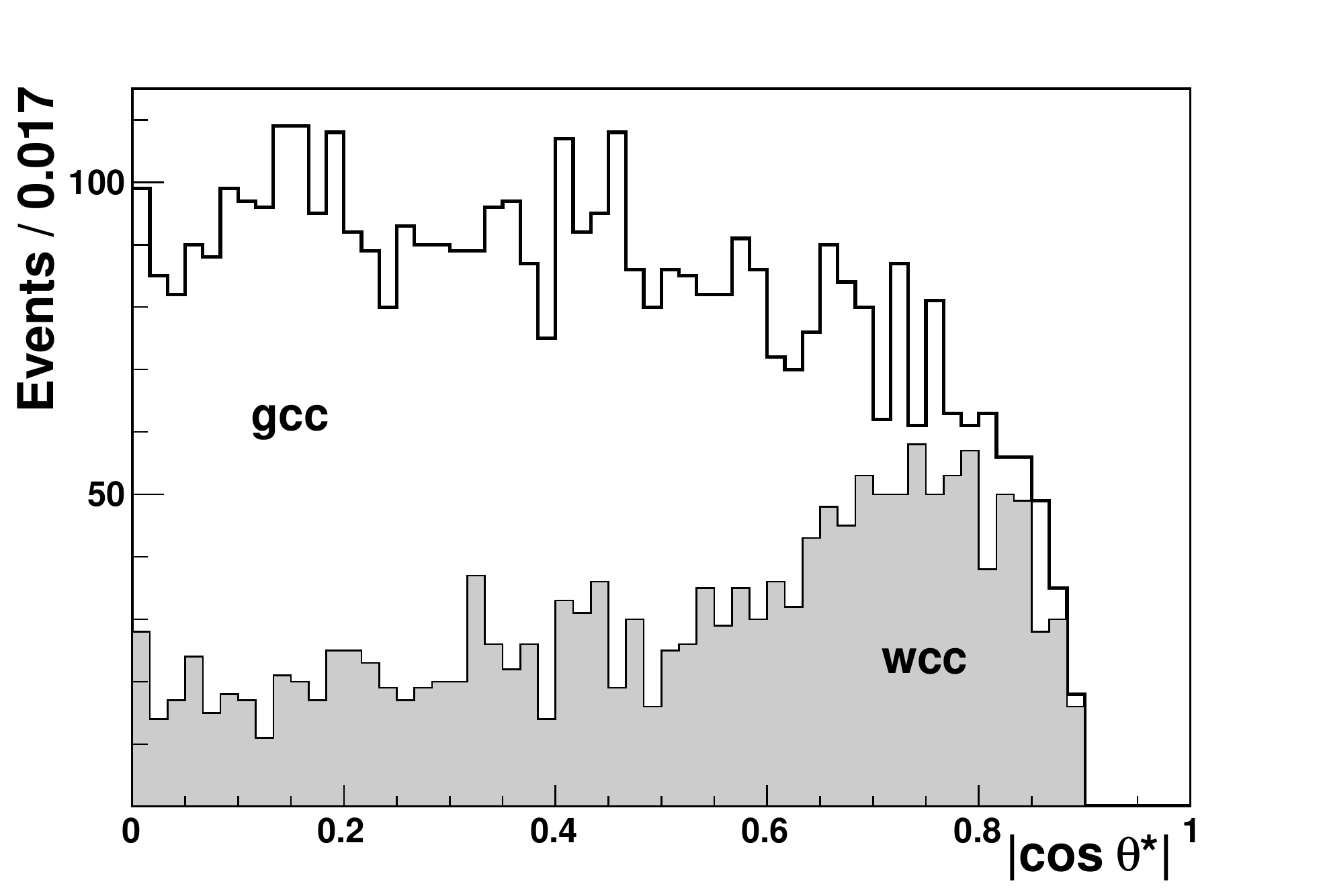}
\end{minipage}
\begin{minipage}[c]{0.46\linewidth}
\includegraphics[totalheight=6.3cm,width=0.96\textwidth]{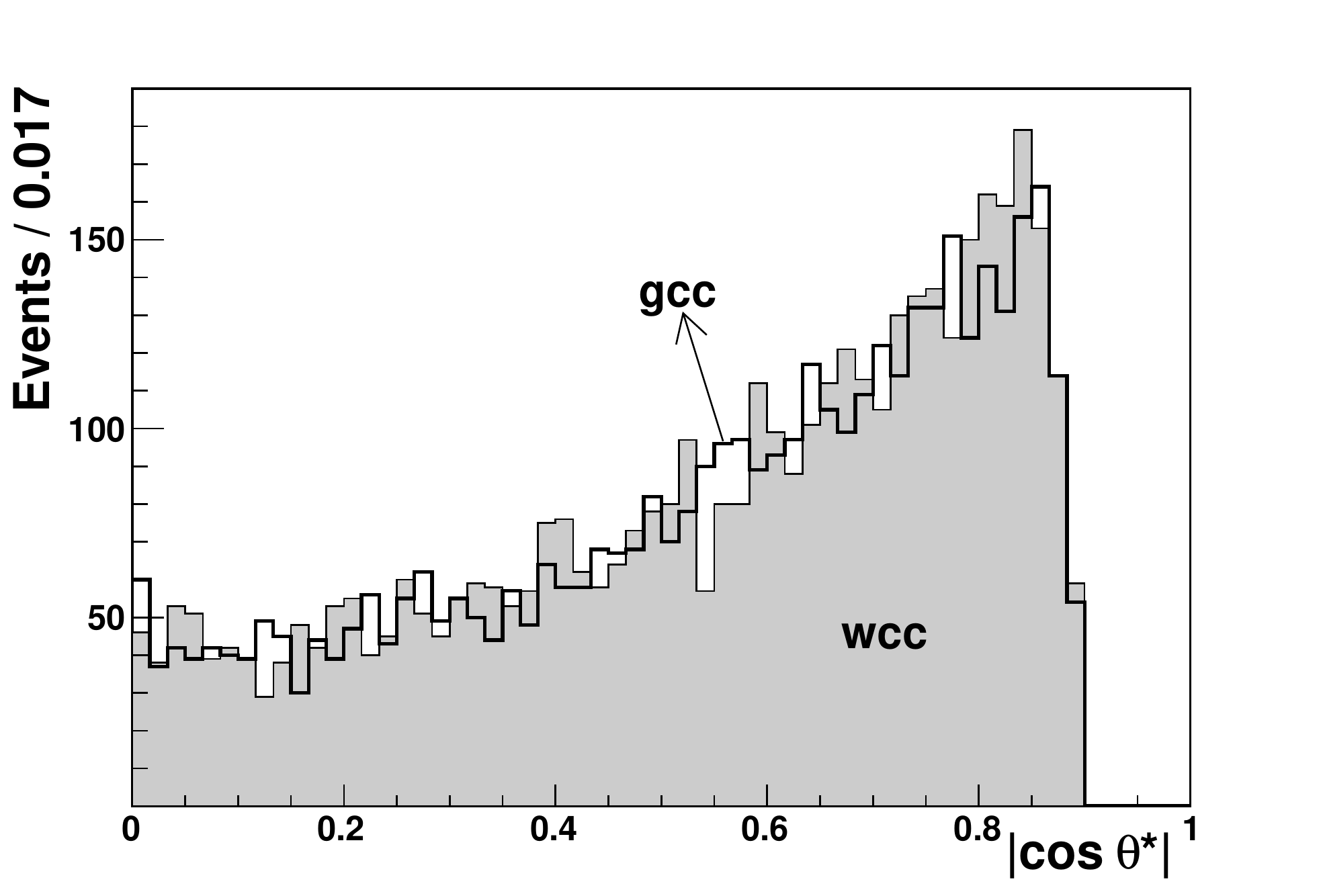}
\end{minipage}
\end{center}
\label{fig:Kaon_angle}
\vspace{-0.3 cm}
\caption{Example of the $gcc$ and $wcc$ distributions of $|\textrm{cos}\ \theta^{*}|$ in the $D^{0}$ centre-of-mass. Left / Right: region of the $D^{0}$ signal / side bands.}
\end{figure}

\vspace{0.2 cm}

\noindent The purity $[s/(s+b)]_{\textrm{NN}}$ is obtained from a simple function applied to the NN output (see Ref. \cite{Celso}). Thereafter, the mass dependence is added as a correction from a fit to the $D^{0}$ mass spectra in bins of $[s/(s+b)]_{\textrm{NN}}$. By respecting the correct $D^{0}$ kinematic dependences, this parametrisation allows us to use $s/(s+b)$ inside $\omega_S$ in an unbiased way.   

\section{Results and Conclusions}

\noindent The experimental results on the gluon polarisation are:

\begin{eqnarray}
\left \langle \frac{\Delta g}{g} \right \rangle^{\textrm{LO}} \ \ &=& -0.08 \pm 0.21 (\textrm{stat}) \pm 0.09 (\textrm{syst}) \ \ @ \langle x_{g} \rangle = 0.11^{+0.11}_{-0.05}.\\
\left \langle \frac{\Delta g}{g} \right \rangle^{\textrm{NLO}} &=& -0.20 \pm 0.21 (\textrm{stat}) \pm 0.09 (\textrm{syst}) \ \ @ \langle x_{g} \rangle = 0.28^{+0.19}_{-0.10}. 
\end{eqnarray}

\vspace{0.3 cm}

\begin{figure}[!h]
\begin{center}
\begin{minipage}[c]{0.46\linewidth}
\includegraphics[totalheight=7.8cm,width=1.02\textwidth]{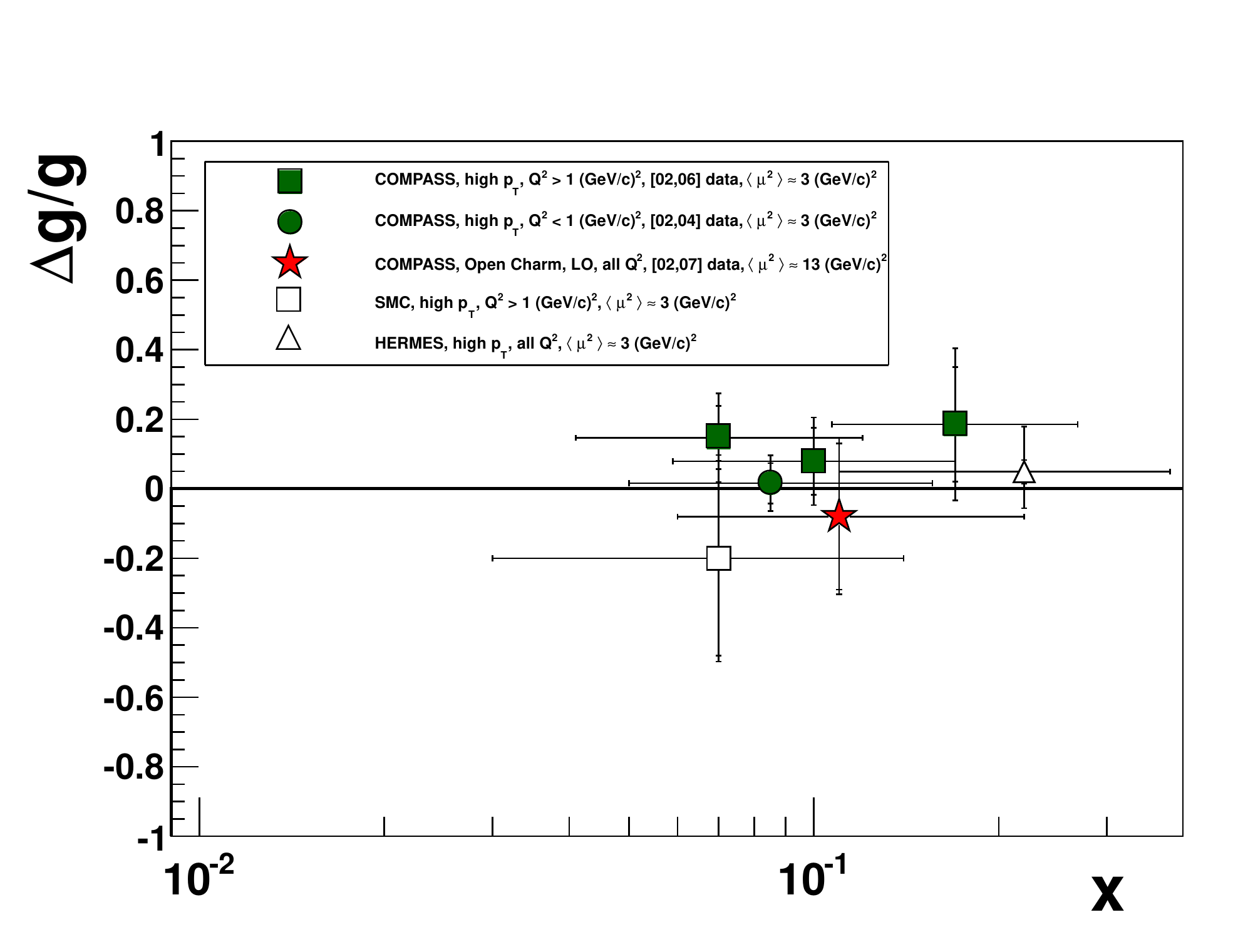}
\end{minipage}
\begin{minipage}[c]{0.46\linewidth}
\includegraphics[totalheight=7.8cm,width=1.02\textwidth]{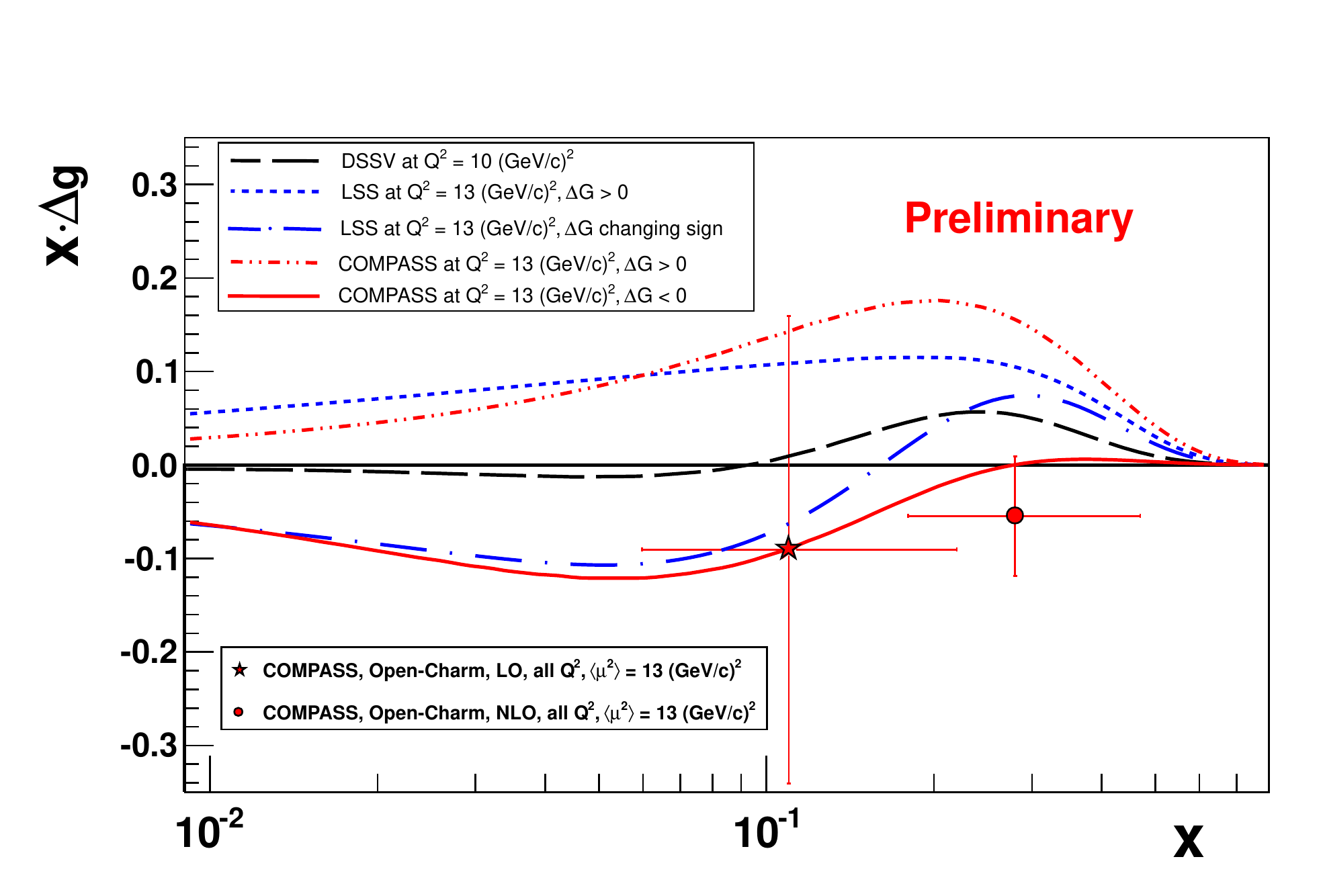}
\end{minipage}
\end{center}
\vspace{-0.2 cm}
\caption{Left: world results on $\Delta g/g$ from LO analyses of the PGF process. Right: parametrisations of $x\Delta g(x,Q^{2})$ together with the LO and NLO results obtained from the open charm analysis. The experimental points of the right plot are obtained from the well known unpolarised gluon distribution $g$, i.e. the $\Delta g/g$ results are multiplied by $g$ at the $x$ points of the measurements.}
\label{fig:DGG}
\end{figure}

\noindent Both results are obtained at a scale of $\langle \mu^{2} \ \rangle$ = 13 (GeV/c)$^{2}$. The LO value is compatible with all the world measurements of the gluon polarisation, as shown in the left plot of Fig. 10. From this figure one can conclude that the gluon polarisation is small and compatible with zero within the range $0.07 < x < 0.20$. Since the method used for the extraction of $\Delta g/g$ at NLO is currently being improved, the result shown in Eq. (11) is still preliminary (the final result might be slightly different).  Nevertheless, the presented value corresponds to the first world measurement of $\Delta g/g$ at NLO.\\
\indent In the right plot of Fig.~\ref{fig:DGG}, the two open charm results for $x\Delta g$ 
are compared to the global fits: the experimental measurements favour small values of the 
gluon contribution to the nucleon spin, i.e. $\Delta G$. Note that the positive solution\footnote[4]{The solutions obtained from the global QCD fits still have large error bands due to the narrow range in $Q^{2}$ which is available from polarised DIS experiments. For example, COMPASS is not able to distinguish a positive solution of $\Delta G$ from a negative one (they have comparable probabilities).} of COMPASS for $\Delta G$ ($\int_{0}^{1}g(x)dx = 0.27 \pm 0.09$) is approximately 3$\sigma$ away from the NLO point. In conclusion, the {\bfseries spin puzzle} of the nucleon still remains to be solved.


\vspace{-0.3 cm}

\end{document}